\documentclass[sigconf]{acmart}
\AtBeginDocument{%
  }
    
\usepackage[most]{tcolorbox}
\usepackage{alltt}
\usepackage{hyperref}
\usepackage{caption}
\usepackage{dsfont}
\usepackage{bbm}
\usepackage{amsfonts}  % provides \mathbb
\usepackage{enumitem}
\usepackage{tabularx}
\usepackage{array}
\usepackage{multirow}
\usepackage{xcolor}
\usepackage{colortbl}
\usepackage{tikz}
\usepackage{booktabs}
\usepackage{arydshln}
\usepackage{ragged2e}   % 用于左对齐示例列
\usepackage{float}
\newcolumntype{C}[1]{>{\centering\arraybackslash}m{#1}}
\newcolumntype{L}[1]{>{\RaggedRight\arraybackslash}m{#1}}

\definecolor{myred}{rgb}{0.7, 0.3, 0.0}
\definecolor{myblue}{HTML}{0a41b8}
\definecolor{mygreen}{HTML}{056b34}
\definecolor{mypurple}{HTML}{5d1e8b}

\setcopyright{acmlicensed}
\copyrightyear{2018}
\acmYear{2018}
\acmDOI{XXXXXXX.XXXXXXX}
\acmConference[Conference acronym 'XX]{Make sure to enter the correct
  conference title from your rights confirmation email}{June 03--05,
  2018}{Woodstock, NY}
\acmISBN{978-1-4503-XXXX-X/2018/06}

\newcommand{\ours}{\texttt{ChatShopBuddy}}

\begin{document}

%\title{Got What You Need: Towards Reliable Conversational Shopping Agents via Reinforcement Learning}
\title{ChatShopBuddy: Towards Reliable Conversational Shopping Agents via Reinforcement Learning}

\author{Yiruo Cheng}
\authornote{This work was done when Yiruo Cheng was doing an internship at JD.com.}
\affiliation{
    \department{Gaoling School of Artificial Intelligence}
    \institution{Renmin University of China}
  \city{Beijing}
  \country{China}
}
\email{chengyr@ruc.edu.cn}

\author{Kelong Mao}
\author{Tianhao Li}
\affiliation{
    \department{JD.com}
  \city{Beijing}
  \country{China}
}
\email{maokelong.1@jd.com}

\author{Jiejun Tan}
\author{Ji-Rong Wen}
\author{Zhicheng Dou}
\authornote{Corresponding author.}
\affiliation{
    \department{Gaoling School of Artificial Intelligence}
    \institution{Renmin University of China}
  \city{Beijing}
  \country{China}
}
\email{dou@ruc.edu.cn}

\renewcommand{\shortauthors}{Yiruo Cheng et al.}

\begin{abstract}
Conversational shopping agents represent a critical consumer-facing application of Large Language Model (LLM)-powered agents, yet how to effectively apply post-training Reinforcement Learning (RL) to optimize such agents remains underexplored. This work investigates RL-based optimization for shopping agents in real-world scenarios, where agents must simultaneously satisfy multiple interdependent objectives spanning objective metrics (product correctness), subjective qualities (persuasiveness), outcome rewards (final response quality), and process rewards (tool efficiency). We present a complete methodology to address this challenge. Specifically, we first construct SmartShopBench, a benchmark that captures diverse shopping intents with a hierarchical evaluation that decomposes complex quality requirements into measurable levels. Building on this evaluation framework, we design Hierarchical Reward Modeling (HRM) to structure mixed reward types through conditional gating that reflects their logical dependencies. 
To enable efficient training, we further propose Dynamic Contrastive Policy Optimization (DCPO), which balances response quality with operational efficiency through dynamic trajectory selection based on reward and reasoning length. Extensive experiments demonstrate that our RL-trained agent, namely \ours{}, consistently outperforms larger models relying on generic reasoning, achieving superior stability rather than merely higher peaks. Our work provides valuable guidance for applying RL to real-world conversational agents.
The dataset and code are available at \href{https://github.com/Ariya12138/ChatShopBuddy_open}{this repository}.
\end{abstract}




\maketitle

\section{Introduction}
\begin{figure}[!t]
    \centering
    \includegraphics[width=\linewidth]{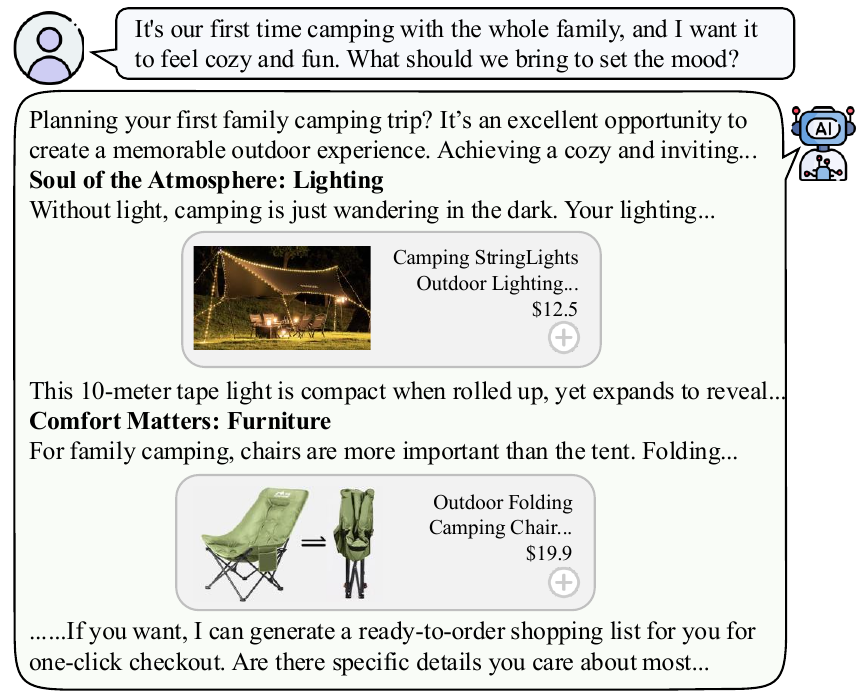}
    \caption{Example of a conversational shopping agent.}
    \label{fig:shoppingagent}
    %\vspace{-5px}
\end{figure}

Conversational shopping agents have emerged as a transformative paradigm for e-commerce, enabling users to express complex, intent-driven queries through natural dialogue rather than keyword-based search~\cite{openai2025shopping,google2025conversational}. Powered by Large Language Models (LLMs), these agents can interpret ambiguous user needs, such as ``It's our first time camping with the whole family, and I want it to feel cozy and fun. What should we bring?'', and synthesize coherent shopping guidance by leveraging external tools, reasoning capabilities, and contextual understanding~\cite{wang_shoppingbench_arxiv,dammu_shopppingagent_wsdm25,tou_shoppingcomp_arxiv}, as shown in Figure~\ref{fig:shoppingagent}. As shopping agents transition from research prototypes to real-world deployment, their ability to provide reliable, complete, and contextually appropriate recommendations becomes critical.

However, conversational shopping agents face significant challenges in real-world deployment. The agent must provide responses that are not only fluent and well-structured but also factually correct, complete, and consistently aligned with user intent. Recent studies reveal persistent and serious issues in recommendation completeness, product relevance, and practical utility~\cite{tou_shoppingcomp_arxiv,shopsimulator,deepshop}. Reinforcement Learning (RL) offers a natural path to effectively address these limitations by directly optimizing agents toward desired behaviors through reward-based training.

Applying RL to conversational shopping agents presents distinct challenges that differentiate it from many RL domains. Unlike mathematical reasoning with definite answers~\cite{shen_satori_icml25,wang_nips}, code generation with testable execution~\cite{wang_codeboost_arxiv,zeng_acl25_acecoder}, or flight bookings with clear success criteria~\cite{tau_bench}, shopping agents must simultaneously optimize across multiple dimensions that resist straightforward verification. A high-quality shopping response requires not only factual correctness in product recommendations but also persuasive presentation, structural coherence, and alignment with subjective user preferences. Furthermore, real-world deployment demands operational efficiency by minimizing reasoning length and reducing unnecessary tool calls to maintain acceptable latency. This combination of multi-dimensional, indirectly verifiable objectives alongside efficiency constraints poses a fundamental challenge for standard RL approaches designed for single-objective or easily verifiable tasks.

In this work, we investigate how RL can be effectively applied to optimize shopping agents toward desired capabilities. We identify multi-dimensional reward modeling, training, and balancing as the central challenge. This encompasses both objective metrics like product correctness and subjective qualities like persuasiveness, as well as directly verified signals such as tool success and indirectly assessed attributes such as structural coherence. To systematically address this challenge, our approach proceeds in three stages.

We first construct \textbf{SmartShopBench}, a benchmark featuring diverse shopping queries and a hierarchical evaluation framework that decomposes complex quality requirements into measurable levels. The Level-1 (L1) Grader validates basic correctness, including product relevance and textual faithfulness, while the Level-2 (L2) Grader assesses higher-order attributes including structural coherence and content depth. 
Building on this evaluation framework, we propose \textbf{Hierarchical Reward Modeling (HRM)}, which synthesizes multi-dimensional objectives into unified training signals through a gated mechanism. HRM explicitly enforces a conditional structure where basic correctness must be satisfied before higher-level quality and efficiency are rewarded, preventing reward hacking and ensuring agents prioritize reliability. 
Finally, we introduce \textbf{Dynamic Contrastive Policy Optimization (DCPO)}, an RL algorithm that balances response quality with operational efficiency. DCPO over-samples trajectories for each query and applies dynamic contrastive selection strategy based on the joint utility of reward scores and reasoning length, explicitly promoting concise, high-quality reasoning paths suitable for deployment.

Extensive experiments on SmartShopBench yield several key findings. 
Firstly, experiments show that \ours{}, an RL-trained model specifically aligned to the task, consistently outperforms larger models relying on generic reasoning, demonstrating that targeted post-training is more effective than scale alone. Secondly, RL primarily improves response stability and consistency rather than merely boosting peak performance, addressing a critical gap for production systems. Finally, we find that encouraging extended reasoning alone does not yield consistent gains on domain-specific tasks, highlighting the importance of aligning training objectives with task requirements.

Our main contributions are summarized as follows:

(1) We systematically investigate the application of RL to conversational shopping agents and propose \textbf{HRM}, a gated reward mechanism that aligns multiple objectives, including reliability, persuasiveness, and efficiency, enabling the shopping agent to achieve high-level quality while ensuring basic correctness.

(2) We propose \textbf{DCPO}, an efficiency-aware RL algorithm that leverages dynamic contrastive selection strategy to jointly optimize both response quality and reasoning length, thereby effectively reducing overall deployment latency.

(3) Extensive experiments show that our task-aligned RL-trained model \ours{} consistently outperforms larger models relying on generic reasoning, improving stability and efficiency without requiring excessively long reasoning length, and highlighting the importance of targeted post-training optimization.

\section{Related Work}

\subsection{Conversational Shopping Agent}
LLM-powered conversational shopping agents~\cite{openai2025shopping,wang_shoppingbench_arxiv,shopsimulator,deepshop} are increasingly being explored as a new paradigm in e-commerce, as they can interact with users to understand needs, reason over preferences and constraints, and strategically invoke external tools to generate high-quality recommendations with coherent and trustworthy explanations~\cite{dammu_shopppingagent_wsdm25,tou_shoppingcomp_arxiv}.
% Existing studies 主要是研究各种不同的模型作为conversational shopping agent的表现，但是没有太多是在探索conversational shopping agent的后训练，特别是rl
While recent work has demonstrated the strong zero-shot or few-shot capability of foundation models in conversational shopping scenarios~\cite{wang_shoppingbench_arxiv,tou_shoppingcomp_arxiv}, results from ShoppingComp~\cite{tou_shoppingcomp_arxiv} indicate that even strong agents still tend to produce redundant internal reasoning and may generate fluent responses that are factually inconsistent with real product attributes or fail to satisfy user intent. These limitations highlight the need for effective post-training methods~\cite{shopsimulator} to better align the agent with user intents, generate more reliable shopping guidance, and reduce unnecessary reasoning.
Therefore, in this work, we propose to leverage RL to enhance conversational shopping agents in terms of intent alignment, response quality, and reasoning efficiency.

\subsection{Reinforcement Learning for Agent}

Reinforcement Learning (RL)~\cite{PPO,GRPO} has emerged as a powerful paradigm for unlocking the potential of autonomous agents~\cite{wang2024survey,search-o1,webthinker,WebSailor,luo_agent_survey,gou_tora_iclr}. 
By optimizing the agent behavior through reward signals, RL allows agents to enhance their complex reasoning and autonomous tool-use capabilities~\cite{DR_survey,tool_light,ASearcher}.

A line of recent studies has therefore concentrated on settings where rewards are objective and directly verifiable~\cite{tool_star,singh_artist_arxiv,li_torl_arxiv,liu_rltf,wang_coevolving_arxiv,wang_rl4coding_survey}. In these scenarios, rewards can be obtained from clear outcomes—such as ground-truth numerical answers in mathematical problems~\cite{zuo_ttrl_arxiv,shen_satori_icml25,wang_nips} or compiler feedback in code generation~\cite{wang_codeboost_arxiv,zeng_acl25_acecoder,zhu_drive_arxiv2511}. This direct verifiability makes the learning process more straightforward and the performance easier to quantify.
However, many real-world agent applications operate in open-ended environments, where rewards are sparse, subjective, or difficult to verify directly~\cite{wang_InfiMed-ORBIT_arxiv,li_researchqa_arxiv}, posing significant challenges for RL-trained agent optimization.
To broaden this scope, some studies have extended RL-driven agents to open-ended domains by incorporating rubric-based rewards~\cite{huang_arxiv_rlrubric,RAR,jin_MR-RML,liu_openrubrics_arxiv,zhou_rubric_arxiv}. 
In this work, we further explore this direction in a real-world conversational shopping setting by studying reward design and policy optimization, aiming to develop the conversational shopping agents that can satisfy user intent, build trust, and deliver high-quality recommendations.

\subsection{Conversational Recommendation}
Conversational recommendation~\cite{jannach_crssurvery_2021,peng_crssruvey_arxiv25} aims to assist users through multi-turn interactions, allowing them to gradually express preferences and refine their intent via iterative feedback.
Existing conversational recommendation studies~\cite{ravaut_crs_eacl24,yuan2026crs,kim_espresso_aaai25,li_ChatCRS_naacl25,zheng_hypercrs_acl25,Kook_naacl25,du_sapient_naacl25,xu_crs_acl25,tajiri_crs_emnlp25,kawamae_kdd25_crs} predominantly focus on improving recommendation accuracy and maintaining semantic coherence in closed-loop conversational settings. While these aspects are important, they mainly adopt a retrieval-centric perspective that treats the task as selecting relevant items and generating coherent dialogue~\cite{liu_crs_emnlp23}.
However, in realistic scenarios—especially in online shopping—user needs extend beyond accurate item retrieval. Users increasingly expect persuasive explanations, meaningful comparisons, and transparent decision support to make informed choices~\cite{tou_shoppingcomp_arxiv}.
To bridge this gap, we adopt an LLM-powered conversational shopping agent with specialized tools, allowing the system to go beyond pure retrieval and offer transparent, persuasive guidance in real-world shopping scenarios.

\begin{table}[t!]
\centering
\small
\caption{Representative queries in SmartShopBench.}
\label{tab:query_categories}
\begin{tabularx}{\linewidth}{p{0.24\linewidth}X}
\toprule
\textbf{Category} & \textbf{Example Query} \\
\midrule
Search-Fuzzy &
``I'm going to meet my girlfriend's parents for the first time. What would be a suitable gift to bring?'' \\

Search-Multi-Constraint &
``I’m looking for a black fully automatic coffee machine with a built-in grinder and milk frother, around \$3000 budget.'' \\

Search-Bundle  &
``I recently want to start baking at home by myself. Please help me put together a set of beginner-level appliances.'' \\

Search-General &
``Recommend a few mini washing machines suitable for baby clothes.'' \\

QA-Compare  &
``I want to learn about the differences between Estée Lauder Double Wear foundation, Lancôme Teint Idole foundation, and Armani Luminous Silk foundation.'' \\

QA-Consultation  &
``Is a double sink with separate sections worth it?'' \\
\bottomrule
\end{tabularx}
\end{table}
\section{SmartShopBench}

To systematically investigate RL for conversational shopping agents, we need an evaluation framework that measures the multi-dimensional quality attributes identified earlier. We construct \textbf{SmartShopBench}, combining diverse real-world shopping queries with a hierarchical evaluation framework designed to decompose complex quality requirements into measurable components.

\subsection{Dataset Construction}

We build the benchmark through a two-stage approach: first establishing a taxonomy of query types grounded in actual user behavior, then using this taxonomy to guide query generation.

\subsubsection{Query Taxonomy}
\label{sec:query_taxonomy}

Analyzing thousands of real user queries from production platforms, we identify six fundamental query types that capture core challenges for conversational shopping agents. The first four are \textbf{search-oriented}:

\textbf{(1) Search-Fuzzy} queries express broad needs without specifying products (e.g., ``something to keep my plants healthy while traveling''), requiring agents to creatively explore the product space.

\textbf{(2) Search-Multi-Constraint} queries impose multiple simultaneous requirements across brand, price, specifications, and other features (e.g., ``quiet blender under \$100 that crushes ice''), demanding joint constraint satisfaction.

\textbf{(3) Search-Bundle} queries request complementary products for coordinated use (e.g., ``complete home coffee setup''), testing reasoning about compatibility and collective utility.

\textbf{(4) Search-General} represents common product requests requiring accurate intent understanding and relevant recommendations.

The remaining two are \textbf{QA-oriented}:

\textbf{(5) QA-Compare} queries request explicit product comparisons (e.g., ``iPhone 15 vs Samsung Galaxy S24''), requiring clear trade-off articulation across attributes.

\textbf{(6) QA-Consultation} queries seek practical advice for specific scenarios (e.g., ``Is the Sony WH-1000XM5 good for frequent flying?''), demanding contextual reasoning and actionable guidance.

Examples for each category are provided in Table~\ref{tab:query_categories}.

\subsubsection{Generation and Annotation Pipeline}

We collect currently available products from major e-commerce platforms and organize them into ten representative categories (apparel, electronics, home goods, etc.). For each query type, we employ an LLM to generate candidate queries referencing these products, using prompts that specify structural and semantic requirements. All queries undergo human review to ensure they are natural, unambiguous, and representative of actual user needs. This yields a final dataset of 1,680 queries across all six categories, of which 1,560 queries are used for training and 120 queries are reserved for testing.

\begin{figure*}
    \centering
    \includegraphics[width=0.9\linewidth]{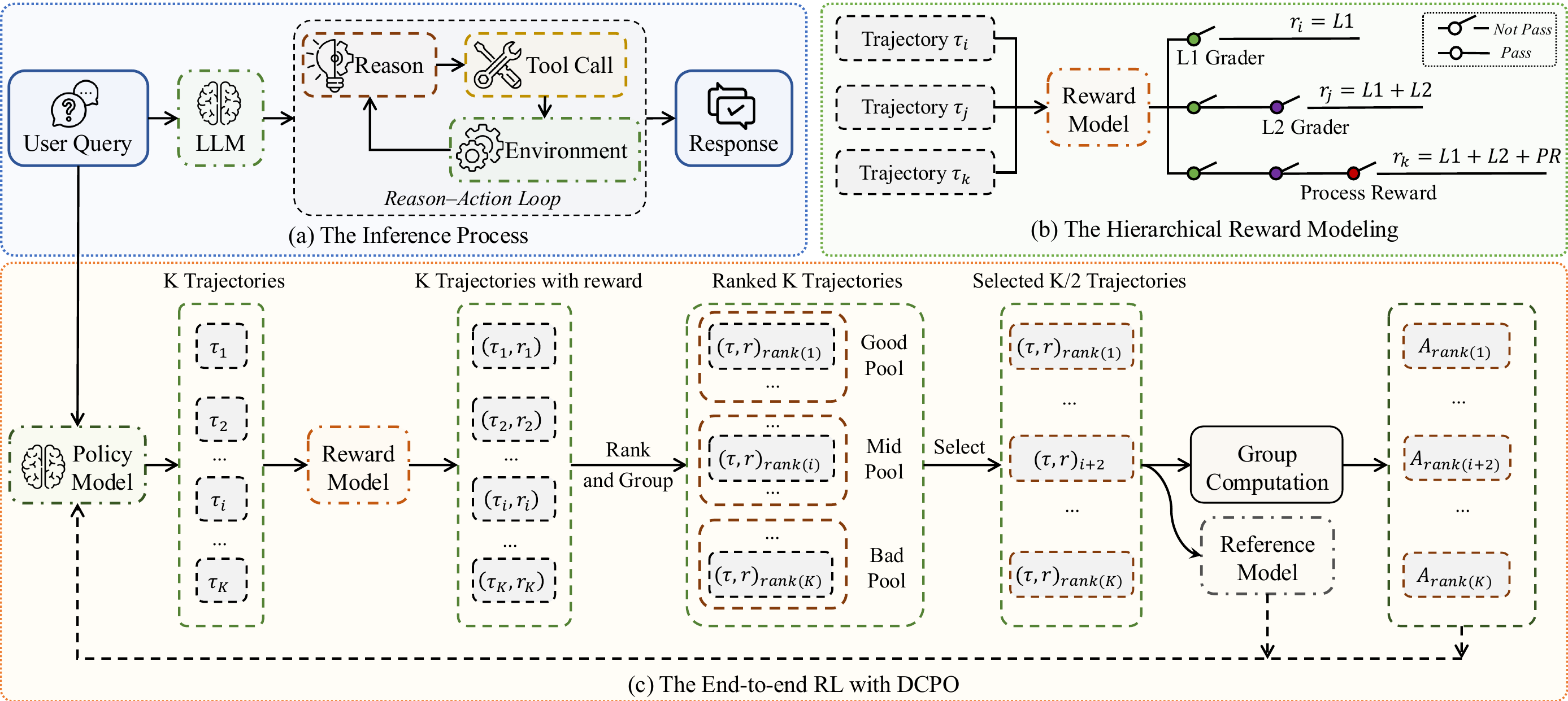}
    \caption{Overview of our conversational shopping agent. Figure(a) is the inference process of our shopping agent. Figure(b) is the Hierarchical Reward Modeling. The green gate indicates whether the L1 Grader passes, the purple gate indicates whether the L2 Grader passes, and the red gate denotes the computation of the process reward. Figure(c) is the end-to-end RL training process with DCPO. We use dynamic contrastive selection strategy choose K/2 trajectories from a total of K.}
    \label{fig:overview}
\end{figure*}

\subsection{Hierarchical Evaluation Framework}
\label{sec:hierarchical_evaluation_framework}

Evaluating conversational shopping agents requires assessing qualities ranging from basic correctness to sophisticated presentation. Collapsing all dimensions into one score obscures failure sources, while treating them independently ignores logical dependencies. We propose a \textbf{Hierarchical Evaluation Framework} with two levels: Level-1 (L1) verifies basic correctness and reliability, while Level-2 (L2) assesses higher-order qualities only after L1 passes. This effectively prevents systems from receiving high scores through eloquent but factually flawed responses.

\subsubsection{L1 Grader}
\label{sec:l1_grader}

The L1 Grader serves as a foundational layer, assessing whether the agent fulfills the user's informational and task-oriented requirements. Conversational shopping agents generate responses consisting of both natural language text and structured product recommendations (displayed as UI cards). The L1 Grader evaluates correctness by analyzing relationships between three core components: the user query, the generated text, and the recommended products. This assessment operates through the following three complementary dimensions:

\textbf{(1) Product Correctness} evaluates whether recommended products satisfy user requirements and are presented appropriately. This dimension includes four checks: 1) product relevance verifies each item falls within the scope of the user request, 2) UI format validation ensures proper structured presentation, 3) UI trigger assessment confirms products are recommended when appropriate and avoided otherwise, and 4) UI completeness checks that all products mentioned in the text have corresponding product cards.

\textbf{(2) Text Relevance} assesses whether the textual response addresses the user query, checking for tangentiality where the agent discusses related but off-topic content, and validating if the response provides coherent guidance connected to the user's needs.

\textbf{(3) Description Faithfulness} verifies that product descriptions in the text accurately and faithfully reflect actual product attributes, guarding against hallucinated features or misleading characterizations that could damage user trust.

A response passes L1 only when all three dimensions pass their checks. We implement this through hybrid evaluation combining rule-based validation for format and structure with LLM-based assessment for semantic correctness. Detailed grading criteria and explanations are provided in Appendix~\ref{appendix:l1_grader}.

\subsubsection{L2 Grader}
\label{sec:l2_grader}

While L1 ensures basic correctness, high-quality shopping responses must also exhibit strong organization and substantive content. The L2 Grader addresses this through a rubric-based assessment measuring seven metrics along two complementary dimensions: structure and depth.

(1) \textbf{Structure}. The structure dimension examines whether the response is coherent and well-organized. This includes whether the response begins with clear problem framing that establishes context and demonstrates understanding of user needs, maintains logical consistency throughout the middle sections where recommendations or comparisons are presented, and concludes with actionable guidance that helps users make informed decisions.

(2) \textbf{Depth}. The depth dimension assesses the overall substantive content quality. This evaluates whether the response provides meaningful comparative analysis when discussing multiple options, demonstrates thoughtful prioritization when presenting alternatives rather than treating all options as equivalent, offers clear and convincing product-level justification explaining why specific recommendations suit the user's needs, and exhibits awareness of potential trade-offs or risks users should consider.

Each metric is assessed using a detailed rubric defining performance levels with concrete examples, enabling consistent evaluation while capturing nuanced quality differences between competent and exceptional responses. Full metric definitions and scoring rubrics are provided in Appendix~\ref{appendix:l2_grader}.

\begin{figure}
    \centering
    \includegraphics[width=\linewidth]{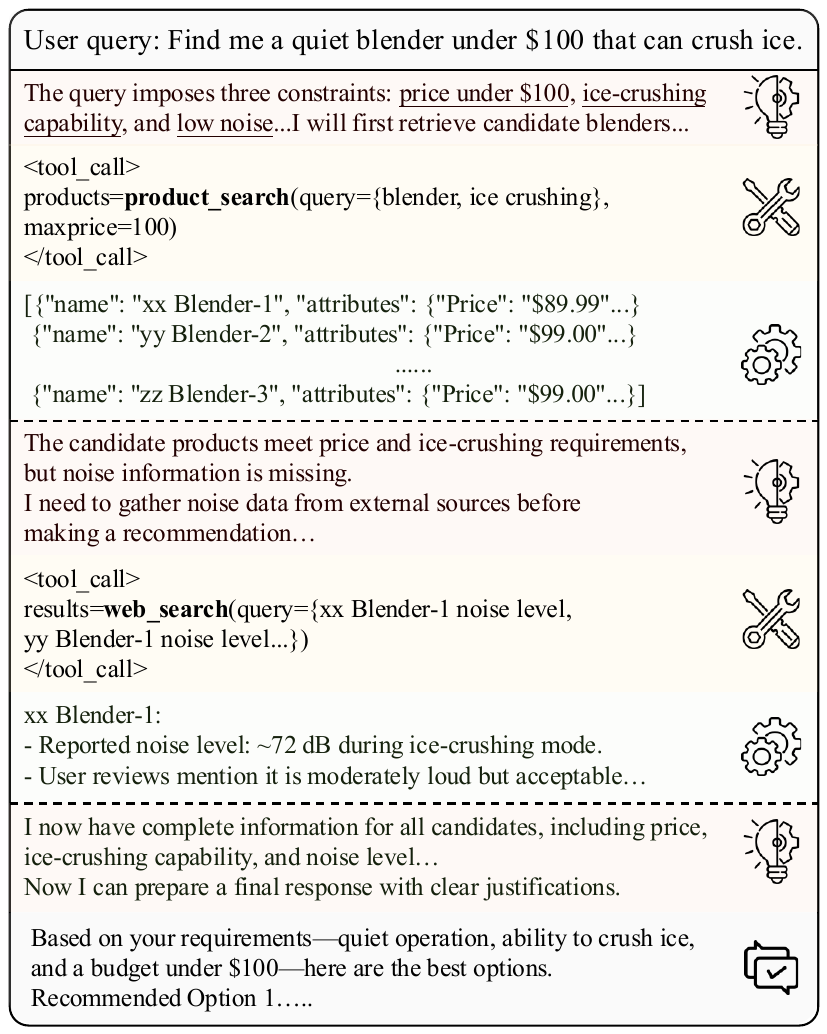}
    \caption{Example trajectory of the shopping agent.}
    \label{fig:trajectory}
    %\vspace{-5px}
\end{figure}

\section{Methodology}

We present our approach \ours{} to training conversational shopping agents through reinforcement learning. We first describe our agent harness that transforms an LLM into a tool-using shopping agent, then introduce HRM to structure multi-dimensional objectives, and finally present DCPO to balance response quality with operational efficiency.

\subsection{Agent Harness}

Our agent harness equips an LLM with external tools and structures its behavior through iterative reasoning and tool invocation. Given a user query $q$, the agent generates a trajectory $\tau$ consisting of multiple reasoning-action-observation cycles before ultimately producing the final response.

\subsubsection{Trajectory Structure.} A trajectory over $T$ steps can be written as $\tau = (R_1, \mathcal{A}_1, \mathcal{O}_1, \ldots, R_T, \mathcal{A}_T, \mathcal{O}_T, y)$. At each step $t$:
\begin{itemize}[leftmargin=1.5em, nosep]
\item $R_t$ is the agent's internal reasoning (chain-of-thought text analyzing the current situation and planning next actions);
\item $\mathcal{A}_t = \{a_t^{(1)}, \ldots, a_t^{(m_t)}\}$ are tool calls issued by the agent;
\item $\mathcal{O}_t = \{o_t^{(1)}, \ldots, o_t^{(m_t)}\}$ are observations returned by the tools.
\end{itemize}

After $T$ steps, the agent generates the final response $y$, which consists of natural language explanation and structured product recommendations. The agent follows a policy $\pi_\theta$ that maps the current context (query and history of reasoning, actions, observations) to the next reasoning and actions.

\subsubsection{Tool Suite.} The agent has access to three tools:
\begin{itemize}[leftmargin=1.5em, nosep]
\item \textsc{web\_search}: retrieves web information for knowledge-intensive questions;
\item \textsc{python\_execute}: executes Python code for computations (e.g., calculating bundle prices, comparing unit costs);
\item \textsc{product\_search}: queries the product database with filters and returns matching products.
\end{itemize}

\subsubsection{Example Agent Trajectory.} Consider the query "Find me a quiet blender under \$100 that can crush ice." The agent begins by reasoning about the constraints ($R_1$: noise level, price, ice-crushing), then calls \textsc{product\_search} with appropriate filters ($\mathcal{A}_1$), receiving candidate products ($\mathcal{O}_1$). Noticing missing noise specifications, it reasons again ($R_2$) and invokes \textsc{web\_search} to look up noise levels ($\mathcal{A}_2$), obtaining supplementary data ($\mathcal{O}_2$). With complete information, it performs final reasoning ($R_3$) and generates the response $y$ containing recommendations with justifications and product cards. Figure~\ref{fig:trajectory} illustrates this process.

Our learning objective is to optimize the policy $\pi_\theta$ to maximize expected trajectory reward:
\begin{equation}
\max_{\theta} \; \mathbb{E}_{\tau \sim \pi_\theta} \bigl[ r(\tau) \bigr].
\end{equation}

\subsection{HRM: Hierarchical Reward Modeling}

Shopping agents must satisfy multiple objectives with natural dependencies: basic correctness is prerequisite to higher-level quality, and efficient tool usage only matters when response quality is satisfactory. We structure rewards through HRM to reflect these logical dependencies.
The total reward decomposes into outcome reward $r_{\text{out}}(\tau)$ evaluating the final response, and process reward $r_{\text{proc}}(\tau)$ evaluating tool usage:
\begin{equation}
r(\tau) = r_{\text{out}}(\tau) + \beta \cdot r_{\text{proc}}(\tau),
\end{equation}
where $\beta$ controls the process reward contribution. Both components are computed hierarchically.

\subsubsection{Hierarchical Outcome Reward}

The outcome reward builds on our hierarchical evaluation framework (Section~\ref{sec:hierarchical_evaluation_framework}). The L1 Grader assesses three fundamental dimensions (product correctness, text relevance, description faithfulness) via binary indicators $c_i(\tau) \in \{0,1\}$:
%\begin{equation}
$G_{\text{L1}}(\tau) = \prod_{i=1}^{3} c_i(\tau)$,
%\end{equation}
which equals 1 only when all dimensions pass. We treat $G_{\text{L1}}$ as a hard feasibility gate: responses with $G_{\text{L1}}(\tau) = 0$ receive zero reward. For feasible responses, the L2 Grader provides a fine-grained quality score $G_{\text{L2}}(\tau) \in [0,1]$ measuring structural coherence and content depth. The final outcome reward is calculated as:
\begin{equation}
r_{\text{out}}(\tau) =
\begin{cases}
1 + \alpha \cdot (G_{\text{L2}}(\tau))^k & \text{if } G_{\text{L1}}(\tau) = 1, \\
0 & \text{if } G_{\text{L1}}(\tau) = 0,
\end{cases}
\end{equation}
where $\alpha$ and $k$ are hyperparameters. The additive constant creates an implicit curriculum prioritizing feasibility before quality, while the quintic transformation $(G_{\text{L2}})^k$ sharpens top-end rewards, preventing the policy from settling at the feasibility boundary.

\subsubsection{Hierarchical Process Reward}

The process reward evaluates tool usage accuracy and efficiency, computed conditionally only for trajectories passing L1 and exceeding L2 quality threshold $\eta$:
\begin{equation}
r_{\text{proc}}(\tau) =
\begin{cases}
\text{Score}(\mathcal{A}, \mathcal{O}) & \text{if } G_{\text{L1}}(\tau) = 1 \text{ and } G_{\text{L2}}(\tau) \geq \eta, \\
0 & \text{otherwise},
\end{cases}
\end{equation}
where $\text{Score}(\mathcal{A}, \mathcal{O})$ is computed by an LLM-based judge evaluating whether tool calls use correct parameters and whether observations effectively contribute to the final response. This gating ensures the agent prioritizes response quality before being rewarded for tool efficiency. Figure~\ref{fig:overview}(b) illustrates this hierarchical structure.

\begin{table*}[!t]
\centering
\caption{Main results on our SmartShopBench. 
For the L1 Grader, we report Product Correctness (PC), Text Relevance (TR), Description Faithfulness (DF), Avg@4, and Pass\^{}4. PR, TR, and DF are averaged over four runs. Avg@4 is the average success rate across four runs, counting a run as successful only if all three dimensions are passed, while Pass\^{}4 is the fraction of responses that pass all three dimensions in all four runs. For the L2 Grader, we report both the mean and standard deviation over four runs.
\# Tokens denotes the number of reasoning tokens.
For open-source models, the best results are in \textbf{bold} and the second are \underline{underlined}. Results from closed-sourced models are in \textcolor{gray!135}{gray} color for reference.
}
%Results from larger or closed-sourced models are in \textcolor{gray!135}{gray} color for reference.}
\label{tab:main_result_performance}
\setlength\tabcolsep{6.8pt}
%\fontsize{8.6pt}{10.4pt}\selectfont
\begin{tabular}{lcccccccc}
\toprule
\multirow{2}{*}{\textbf{Model}}
& \multirow{2}{*}{\textbf{\# Tokens}}
& \multicolumn{5}{c}{\textbf{L1 Grader}} &  \multicolumn{2}{c}{\textbf{L2 Grader}}\\

\cmidrule(lr){3-7} \cmidrule(lr){8-9}
& & \textbf{PC} & \textbf{TR} & \textbf{DF} & \textbf{Avg@4}  & \textbf{Pass\^{}4}  & \textbf{Avg@4} & \textbf{Std@4}
    \\
  
% \cmidrule(lr){3-5} \cmidrule(lr){6-8} \cmidrule(lr){9-11} \cmidrule(lr){9-10} 
%  &  & Success & Path 
%    & Success & Path 
%    & Success & Path 
%    & Success & Path 
%    & Correct & Path \\
\midrule
\rowcolor[HTML]{f0f0f0}
\multicolumn{9}{c}{\textit{\textbf{Non-Thinking Model}}} \\
DeepSeek-V3.2-chat        &  - & 69.80 & 98.75 & 66.90 & 43.33 & 7.50 & 0.4550 & \textbf{0.0058} \\
GLM-4.7(think off) & - & 61.67 & \underline{99.60} & 67.90 & 38.98 & 5.00 & 0.4825 & \underline{0.0096}\\
Kimi-K2.5(think off) & - & 31.45 & 72.70 & 77.90 & 21.65 & 5.00 & 0.2850 & 0.0520\\
Qwen3-235B-A22B-Instruct-2507 & - & 42.30 & 99.40 & 65.42 & 25.20 & 2.50 & 0.5775 & 0.0126 \\
Qwen3-30B-A3B-Instruct-2507 & - & 14.57 & 93.97 & 42.30 & 3.92 & 0.00 & 0.5625 & 0.0299\\

GPT-5.2(think off)    & \textcolor{gray!135}{-} & \textcolor{gray!135}{80.03} & \textcolor{gray!135}{99.80} & \textcolor{gray!135}{91.88} & \textcolor{gray!135}{66.88} & \textcolor{gray!135}{35.00} & \textcolor{gray!135}{0.6925} & \textcolor{gray!135}{0.0150} \\
Gemini-3-Flash-Preview(minimal)   & \textcolor{gray!135}{-} & \textcolor{gray!135}{74.80} & \textcolor{gray!135}{100.00} & \textcolor{gray!135}{71.22} & \textcolor{gray!135}{48.75} & \textcolor{gray!135}{9.20} & \textcolor{gray!135}{0.5600} & \textcolor{gray!135}{0.0163}  \\
Doubao-Seed-1.8(think off) & \textcolor{gray!135}{-} & \textcolor{gray!135}{73.53} & \textcolor{gray!135}{98.75} & \textcolor{gray!135}{49.38} & \textcolor{gray!135}{30.20} & \textcolor{gray!135}{1.70} & \textcolor{gray!135}{0.3525} & \textcolor{gray!135}{0.0222}\\

\midrule
\rowcolor[HTML]{f0f0f0}
\multicolumn{9}{c}{\textit{\textbf{Thinking Model}}} \\
DeepSeek-V3.2-reasoner & 1,412 & 86.05 & \textbf{99.80} & \underline{79.38} & \underline{62.10} & \underline{19.20} & 0.6200 & 0.0271 \\
GLM-4.7 & 764 & 64.80 & \underline{99.60} & 65.40 & 37.50 & 2.50 & 0.4250 & 0.0100 \\
Kimi-K2.5 & 515 & 70.42 & 98.95 & 77.72 & 47.73 & 12.50 & 0.4900 & 0.0115 \\
%Qwen3-Max-Thinking & \\
Qwen3-235B-A22B-Thinking-2507 & 1,526 & 36.88 & 99.40 & 24.80 & 8.32 & 0.00 & \textbf{0.7015} & 0.0150 \\
Qwen3-30B-A3B-Thinking-2507 &  2,803 & 47.30 & 98.55 & 31.70 & 12.70 & 0.00 & 0.6275 & \underline{0.0096} \\
GPT-5.2 & \textcolor{gray!135}{N/A} & \textcolor{gray!135}{88.15} & \textcolor{gray!135}{99.80} & \textcolor{gray!135}{99.20} & \textcolor{gray!135}{75.85} & \textcolor{gray!135}{42.50} & \textcolor{gray!135}{0.6750} & \textcolor{gray!135}{0.0058}\\
%Claude-Opus-4.5 & \\
Gemini-3-Pro-Preview & \textcolor{gray!135}{N/A} & \textcolor{gray!135}{80.40} & \textcolor{gray!135}{99.80} & \textcolor{gray!135}{70.85} & \textcolor{gray!135}{55.03} & \textcolor{gray!135}{15.80} & \textcolor{gray!135}{0.6725} & \textcolor{gray!135}{0.0050}\\
Doubao-seed-1.8 & \textcolor{gray!135}{1,056} & \textcolor{gray!135}{76.47} & \textcolor{gray!135}{99.38} & \textcolor{gray!135}{58.75} & \textcolor{gray!135}{39.77} & \textcolor{gray!135}{3.30} & \textcolor{gray!135}{0.2825} & \textcolor{gray!135}{0.0189}\\
\rowcolor[HTML]{ecf0ff}
\ours{}-SFT & 1,200 & \underline{86.88} & 98.55 & 72.28 & 60.40 & 18.30 &  0.4800 & 0.0606  \\
\rowcolor[HTML]{ecf0ff}
\ours{}-SFT-RL   & 633 & \textbf{93.35}  & \underline{99.60} & \textbf{84.97} & \textbf{75.22} & \textbf{34.20} & \underline{0.6325} & \underline{0.0096}\\
\bottomrule
\end{tabular}
\end{table*}

\subsection{Dynamic Contrastive Policy Optimization}

Real-world deployment requires minimizing reasoning latency, which we quantify through reasoning length $L(\tau)$—the total token count of internal reasoning. DCPO explicitly optimizes the quality-efficiency trade-off through dynamic trajectory selection.

For each query $q$, we sample $K$ candidate trajectories and rank them lexicographically: first by reward $r(\tau)$ descending, then by reasoning length $L(\tau)$ ascending for ties. This prioritizes high-reward trajectories while preferring concise reasoning when quality is comparable. Based on this, we partition trajectories into three equal-sized pools $\mathcal{P}_{\text{good}}$, $\mathcal{P}_{\text{mid}}$, $\mathcal{P}_{\text{bad}}$ (top, middle, bottom). We construct a selected subset $\mathcal{S}(q)$ of approximately $K/2$ trajectories via:

\begin{itemize}[leftmargin=1.5em, nosep]
\item \textbf{Anchors}: Select $\tau_{\text{best}}$ (top-ranked from $\mathcal{P}_{\text{good}}$) as positive reference and $\tau_{\text{worst}}$ (bottom-ranked from $\mathcal{P}_{\text{bad}}$) as negative reference.
\item \textbf{Stratified Sampling}: Fill remaining $K/2 - 2$ slots with random, proportionally representative samples from all pools.
\end{itemize}
We compute advantages over $\mathcal{S}(q)$ using reward normalization:
\begin{equation}
\hat{A}(\tau) = \frac{r(\tau) - \mu(q)}{\sigma(q) + \delta},  
\end{equation}
where $\delta$ ensures numerical stability, $ \mu(q) = \frac{1}{|\mathcal{S}(q)|} \sum_{\tau \in \mathcal{S}(q)} r(\tau)$ and $\sigma(q) = \sqrt{\frac{1}{|\mathcal{S}(q)|} \sum_{\tau \in \mathcal{S}(q)} (r(\tau) - \mu(q))^2}$.  The policy optimization objective is:
\begin{equation*}
\begin{aligned}
\mathcal{J}(\theta) 
&= \mathbb{E}_{\tau \sim \mathcal{S}(q)} \Big[ 
    \min \Big(
        \rho_\theta(\tau) \hat{A}(\tau),
        \text{clip}(\rho_\theta(\tau), 1-\epsilon, 1+\epsilon) \hat{A}(\tau)
    \Big)
\Big] \\
&\quad - \lambda \, \mathbb{D}_{\text{KL}}(\pi_\theta \| \pi_{\text{ref}}),
\end{aligned}
\end{equation*}
where $\rho_\theta(\tau) = \frac{\pi_\theta(\tau)}{\pi_{\text{old}}(\tau)}$ is the importance sampling ratio, $\epsilon$ is the clipping threshold, and $\lambda$ controls the KL penalty. Figure~\ref{fig:overview}(c) illustrates the DCPO training process.

\section{Experimental Settings}

\subsection{Dataset and Evaluation Metrics}

\subsubsection{Dataset}
We use SmartShopBench for both training and evaluation. The benchmark contains 1,680 queries across six distinct categories (Section~\ref{sec:query_taxonomy}): Search-Fuzzy, Search-Multi-Constraint, Search-Bundle, Search-General, QA-Compare, and QA-Consultation. Among these, 1,560 queries are used for training, while the remaining 120 queries are reserved for testing.

\subsubsection{Evaluation Protocol}
We adopt the Hierarchical Evaluation Framework from Section~\ref{sec:hierarchical_evaluation_framework}. Each response is evaluated in four independent runs to account for sampling variance.

For L1 evaluation, we assess three dimensions (product correctness, text relevance, description faithfulness) using rule-based checks and LLM evaluation. A run succeeds only if all three dimensions pass. We report: (1) per-dimension average pass rate across four runs, (2) \textbf{Avg@4}, the average success rate over four runs, and (3) \textbf{Pass\^{}4}, the fraction of responses passing all four runs.

For L2 evaluation, we report the mean and standard deviation of quality scores across four independent runs, providing a more fine-grained assessment of both structural coherence and content depth beyond binary correctness.

\subsection{Baselines}

We compare against a comprehensive set of baseline models with varying reasoning capabilities. Open-source models include DeepSeek-V3.2~\cite{deepseekv32}, GLM-4.7~\cite{glm4.7}, Kimi-K2.5~\cite{kimik25}, and Qwen3 series (235B-A22B, 30B-A3B) with standard and thinking variants~\cite{qwen3_tech_report}. Proprietary models include GPT-5.2~\cite{openai_gpt52}, Gemini-3-Flash-Preview~\cite{google_gemini3_flash}, Gemini-3-Pro-Preview~\cite{google_gemini3_pro}, and Doubao-seed-1.8~\cite{seed1.8_modelcard_2025}. Models with extended reasoning modes (e.g., GPT-5.2) are evaluated both with reasoning enabled and disabled where applicable, allowing us to assess the impact of deliberate reasoning on shopping tasks.

\subsection{Implementations}

We use Qwen3-30B-A3B-Thinking-2507~\cite{qwen3_tech_report} as the base model. The agent is first supervised fine-tuned on 480 high-quality annotated trajectories with batch size 8 and learning rate 7.0e-6 for 2 epochs, then trained with DCPO for 2 epochs using batch size 16 and learning rate 1e-6. During RL training, we sample $K=16$ trajectories per query and select $K/2=8$ via dynamic contrastive selection. Maximum sequence length during training is 32,768 tokens.
For inference, we use maximum context length 81,920 tokens, temperature 1.0, and top-p 0.9. 

For reward computation, both the L1 and L2 graders, as well as the process reward, are based on DeepSeek-V3.2-reasoner~\cite{deepseekv32}. We set $\alpha=0.5$, $\beta=0.05$, $\eta=0.7$, $k=5$ in HRM. 
For the tool implementations, \textsc{web\_search} uses Zhipu's \texttt{search\_pro} API and \textsc{product\_search} is a commercial product search API from real-world industrial deployment. Additional implementation details are provided in Appendix~\ref{appendix:implementation_details}.

\section{Experimental Results}

\subsection{Main Results}
Table~\ref{tab:main_result_performance} presents the main results on SmartShopBench under both the L1 and L2 Graders, leading to several clear observations.

\textbf{(1) Task-aligned Reinforcement Learning Outweighs Model Scale and Generic Reasoning.}
With explicit task-aligned optimization, \ours{} demonstrates strong overall performance, achieving a product correctness of 93.35, description faithfulness of 84.97, Avg@4 of 75.22, and Pass\^{}4 of 34.20.
Compared with the larger DeepSeek-V3.2-reasoner, ShoppingAgent-RL outperforms across nearly all L1 metrics, gaining +7.30 in product correctness, +5.59 in description faithfulness, +13.12 in Avg@4, and +15.00 in Pass\^{}4.
Notably, under the L2 Grader, it attains an Avg@4 of 0.6325—comparable to DeepSeek-V3.2-reasoner (0.6200)—while showing much lower variance (0.0096). This indicates that stable and high-quality conversational shopping performance is primarily driven by task-aligned optimization, rather than model scale or general-purpose reasoning ability.

\textbf{(2) Reinforcement Learning Improves Stability beyond Response Quality.}
Compared with supervised fine-tuning, the RL-optimized \ours{} achieves substantial improvements in both task success and consistency. Specifically, Pass\^{}4—the fraction of responses that pass all L1 dimensions in all four runs—rises from 18.30 to 34.20, demonstrating that RL significantly increases the likelihood of fully reliable responses across multiple trials. Meanwhile, the average L2 performance improves significantly (from 0.4800 to 0.6325), and the standard deviation over four runs drops sharply from 0.0606 to 0.0096. This indicates that RL’s main benefit is not just raising peak response quality but ensuring consistent, reliable outputs across runs, which are critical for practical deployment.

\textbf{(3) Extended Reasoning Alone Does Not Always Translate into Gains in Domain-Specific Tasks.}
While thinking-enabled models generally achieve higher response quality on SmartShopBench, the benefits of thinking are not consistent across both L1 and L2 evaluations.
For example, DeepSeek-V3.2-reasoner (1,412 reasoning tokens) demonstrates clear gains from thinking, with Pass\^{}4 improving by 19.20 and L2 Avg@4 reaching 0.6200. In contrast, GLM-4.7 (764 reasoning tokens) shows a slight decline in L1 Pass\^{}4 by 2.50, despite achieving a comparable L2 Avg@4 of 0.4250. Similarly, Qwen3-235B-A22B-Thinking-2507 (1,526 reasoning tokens) achieves a high L2 Avg@4 of 0.7015, but exhibits unstable task performance, with product correctness at 36.88 and L1 Avg@4 at 8.32.
These results indicate that reasoning does not uniformly improve performance across different graders. Without task-aligned optimization, longer or more complex reasoning can sometimes lead to over-reasoning or reasoning drift, which may hurt task outcomes rather than help.

\begin{figure}[!t]
    \centering
    \includegraphics[width=\linewidth]{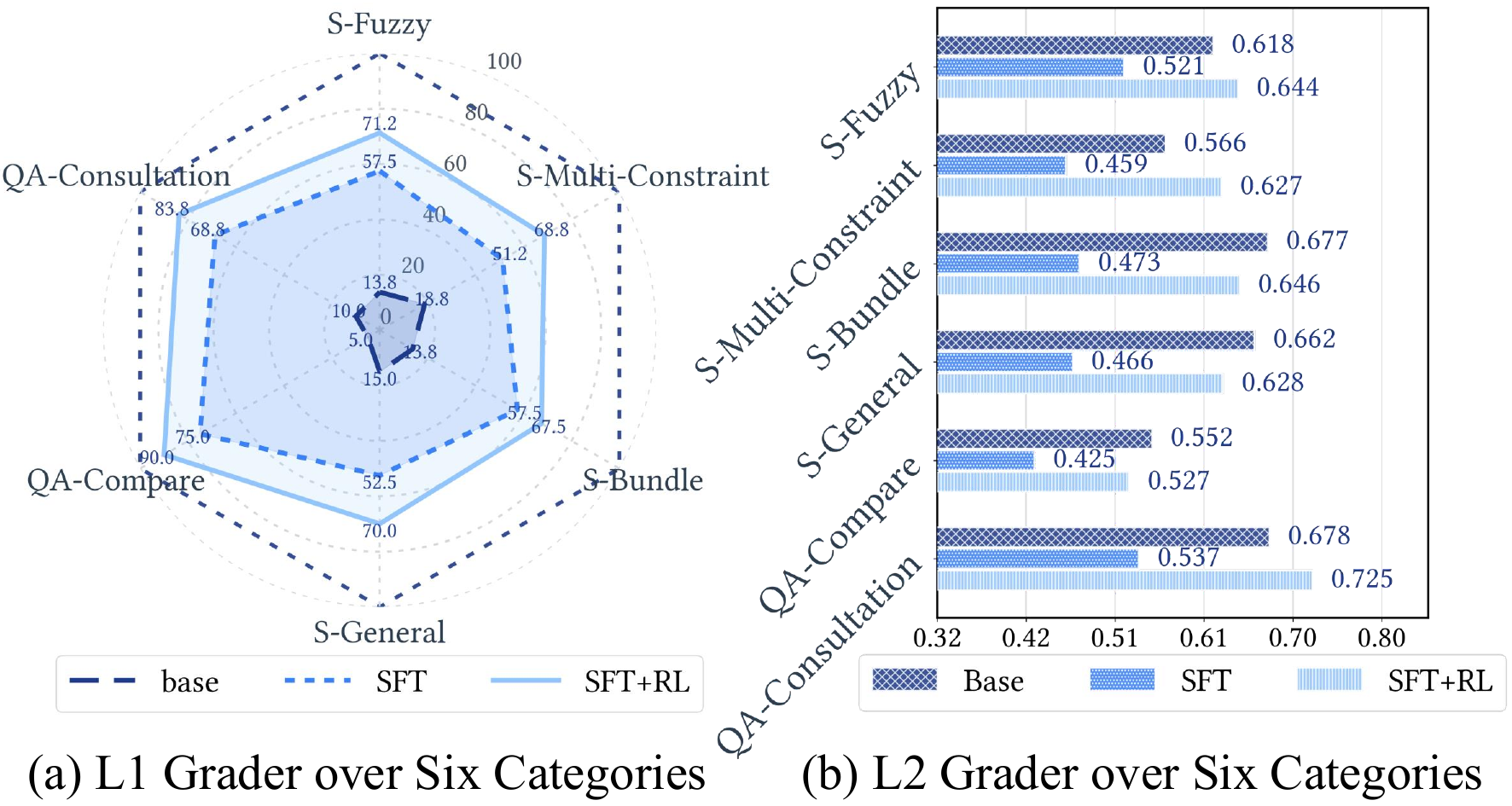}
    \caption{Performance of \ours{} across six shopping categories. We report Avg@4 over four independent runs.}
    \label{fig:gain_six_categories}
\end{figure}

\subsection{Performance across Shopping Categories}

Figure~\ref{fig:gain_six_categories} shows the performance gains from the training process across six distinct shopping intent categories.
Under the L1 Grader, which measures whether the agent fulfills basic user needs, training consistently improves performance across all categories. Supervised fine-tuning (SFT) yields substantial gains over the base model, demonstrating effective adaptation to complex shopping tasks, while RL further enhances performance, with particularly strong improvements in QA-Compare, QA-Consultation, and multi-constraint search intents.

By contrast, the L2 Grader evaluates higher-level qualities such as structural coherence and content depth. Under this stricter metric, SFT alone actually causes a performance drop across all categories, indicating that while it improves basic task capabilities, it does not optimize for higher-level response quality. Reinforcement learning, however, not only recovers this decline but also exceeds the base model in every category, with particularly strong gains in QA-Consultation and QA-Compare.

Overall, these results highlight a clear training dynamic: SFT improves fundamental task success, but may degrade high-level response quality, whereas RL acts as a corrective stage, aligning the model with both basic user needs and more sophisticated output standards, resulting in robust gains across all shopping intents.

\begin{figure}[!t]
    \centering
    \includegraphics[width=0.8\linewidth]{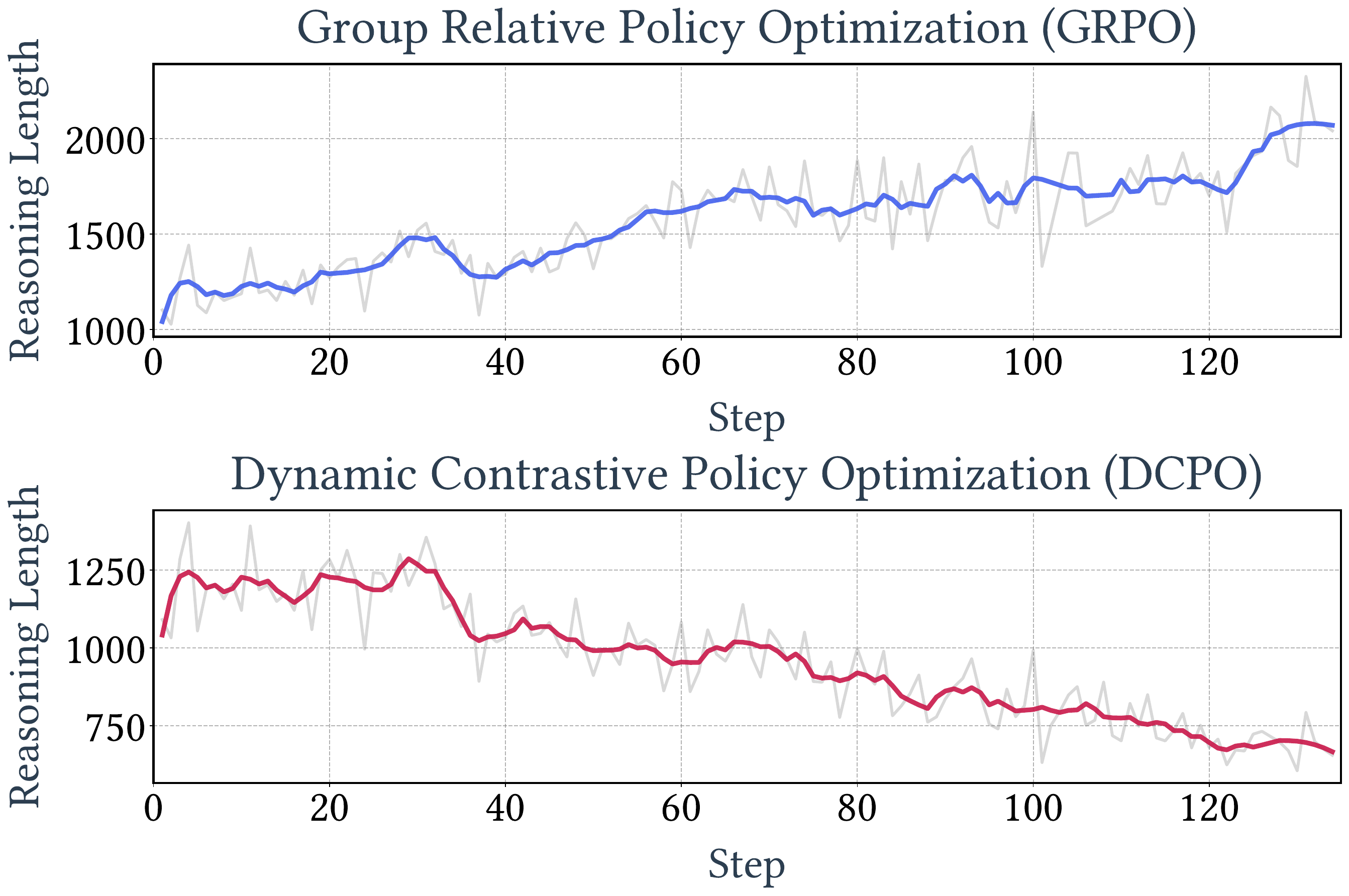}
    \caption{Comparison of reasoning length during the training process DCPO vs. GRPO.}
    \label{fig:reasoning_length}
    %\vspace{-5px}
\end{figure}

\begin{figure}[t]
    \centering
    \includegraphics[width=\linewidth]{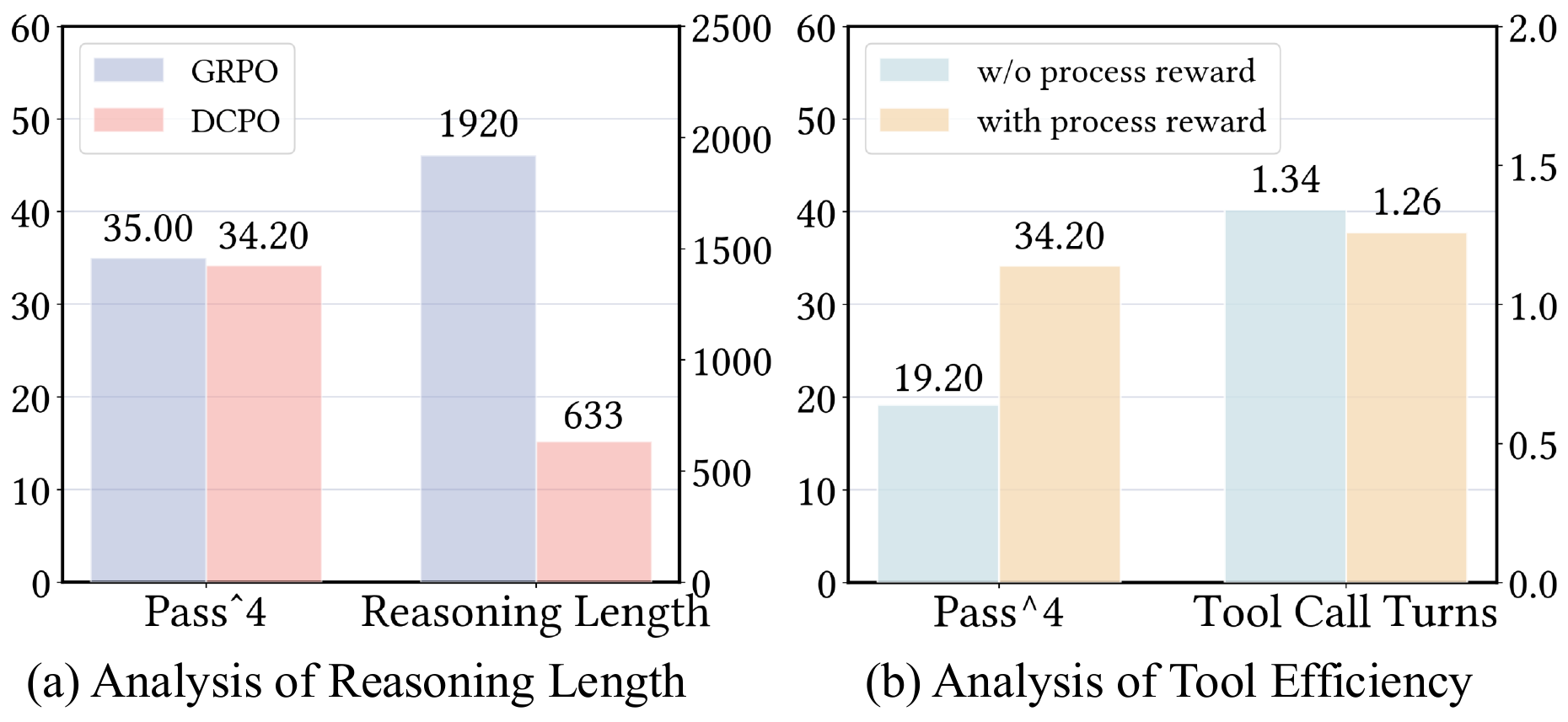}
    \caption{Analysis of Operational Efficiency.}
    \label{fig:analysis_operational_efficiency}
    %\vspace{-5px}
\end{figure}

\subsection{Analysis of Operational Efficiency}

Operational efficiency is crucial for reasoning-enabled shopping agents, as inference latency and tool usage efficiency impact user experience. In this section, we analyze efficiency in terms of reasoning length and tool usage, and assess whether gains can be made without compromising task performance.

\paragraph{Reasoning Length Efficiency.}
Reasoning length directly affects online latency: longer reasoning trajectories increase inference cost without necessarily improving response quality. As shown in Figure~\ref{fig:reasoning_length}, DCPO exhibits a clear reduction in reasoning length throughout training, whereas GRPO progressively produces longer reasoning sequences, reflecting fundamentally different optimization dynamics.
This divergence during training carries over to inference-time behavior. As illustrated in Figure~\ref{fig:analysis_operational_efficiency} (a), models trained with DCPO generate shorter reasoning length at inference while achieving task performance comparable to GRPO. This suggests that DCPO does not rely on increasingly verbose reasoning to obtain high rewards. Instead, by leveraging the dynamic contrastive selection strategy, DCPO encourages efficient and effective reasoning patterns, reducing unnecessary computation and inference latency while preserving strong task success.

\paragraph{Tool Usage Efficiency.}
Tool invocation behavior also impacts operational cost, as excessive or redundant calls increase execution overhead. To examine this effect, we introduce a process reward that explicitly encourages efficient tool usage.
Interestingly, introducing the process reward leads to a substantial improvement in task performance—measured by Pass@4—while simultaneously enhancing tool-call efficiency.
As shown in Figure~\ref{fig:analysis_operational_efficiency}(b), the model achieves significantly higher Pass@4 scores with fewer tool-call turns, suggesting that encouraging more precise tool usage not only improves solution quality but also reduces unnecessary computation and operational overhead.

\subsection{Ablation Studies}
We conduct ablation studies in Table~\ref{tab:ablation} to validate the effectiveness of each component in our training framework.

(1) Importance of DCPO Training:
Removing DCPO training leads to the most significant performance drop across all metrics. Specifically, the L1 Avg@4 decreases from 75.22 to 60.40 and Pass\^{}4 drops from 34.20 to 18.30, accompanied by a substantial decline in L2 Avg@4 and increased variance. This highlights the central role of DCPO training in aligning the model with task-oriented objectives and ensuring stable performance.

(2) Effectiveness of Hierarchical Reward:
Removing the hierarchical reward results in a noticeable degradation in L1 performance, indicating weaker reliability. Although the L2 Avg@4 slightly increases, the overall performance remains inferior to the full model. This suggests that hierarchical reward signals are essential for generating responses that are both reliable and task-aligned.

(3) Contribution of Process Reward:
Eliminating the process reward leads to a clear drop in Pass\^{}4, reflecting reduced consistency across multiple runs. Meanwhile, L2 Avg@4 remains comparable with lower variance, indicating that process-level reward primarily improves multi-step reliability and stability during decision making, rather than directly boosting higher-level response quality.

% Overall, these results demonstrate that different reward components contribute in distinct ways: CAPO ensures core task alignment, hierarchical rewards improve goal-directed optimization, and process rewards enhance consistency and stability, with their integration delivering the best overall performance.

\begin{table}[t]
\centering
\small
\caption{Ablation study of our \ours{}, with the best scores shown in \textbf{bold} and the second-best in \underline{underlined}.}
\label{tab:ablation}
\begin{tabular}{lcccc}
\toprule
\multirow{2}[2]{*}{\textbf{Method}} & \multicolumn{2}{c}{\textbf{L1 Grader}} & \multicolumn{2}{c}{\textbf{L2 Grader}}  \\
\cmidrule(lr){2-3} \cmidrule(lr){4-5}
 & Avg@4 & Pass\^{}4 & Avg@4 & Std@4  \\
\midrule
\rowcolor[HTML]{ecf0ff}
\ours{}  & \textbf{75.22} & \textbf{34.20} & \underline{0.6325} & 0.0096 \\
\midrule
\quad w/o DCPO Training (SFT) & 60.40 & 18.30 & 0.4800 & 0.0606  \\
\quad w/o Hierarchical Reward & 66.45 & 26.90 & \textbf{0.6450} & \underline{0.0058}  \\
\quad w/o Process Reward & 66.05 & 19.20 & \underline{0.6325} & \textbf{0.0050} \\
\bottomrule
\end{tabular}
\end{table}

\section{Conclusion}
We present \ours{}, our task-aligned RL-trained agent for real-world conversational shopping, effectively addressing the challenge of optimizing multiple, indirectly verifiable objectives while ensuring efficiency. We introduce SmartShopBench for hierarchical evaluation, propose Hierarchical Reward Modeling (HRM) to integrate correctness, persuasiveness, and efficiency through conditional gating that reflects their logical dependencies, and develop Dynamic Contrastive Policy Optimization (DCPO) to dynamically select concise, high-quality reasoning trajectories that balance response quality and operational efficiency. Extensive experiments show that \ours{} consistently outperforms larger models relying on generic reasoning, achieving greater stability, efficiency, and overall user satisfaction. Our work provides practical guidance for designing reliable, persuasive, and efficient conversational shopping agents ready for real-world deployment.

% \begin{acks}
% \end{acks}

\balance
\bibliographystyle{ACM-Reference-Format}
\bibliography{mybib}


\newpage
\section*{Appendix}
\appendix

\section{Hierarchical Evaluation Framework}
\subsection{L1 Grader}
\label{appendix:l1_grader}
The Level-1 (L1) Grader evaluates responses along three dimensions: Product Correctness, Text Relevance, and Description Faithfulness. The prompt are shown in Figure~\ref{l1_prompt}.

\begin{figure*}
    \centering
    \includegraphics[width=\linewidth]{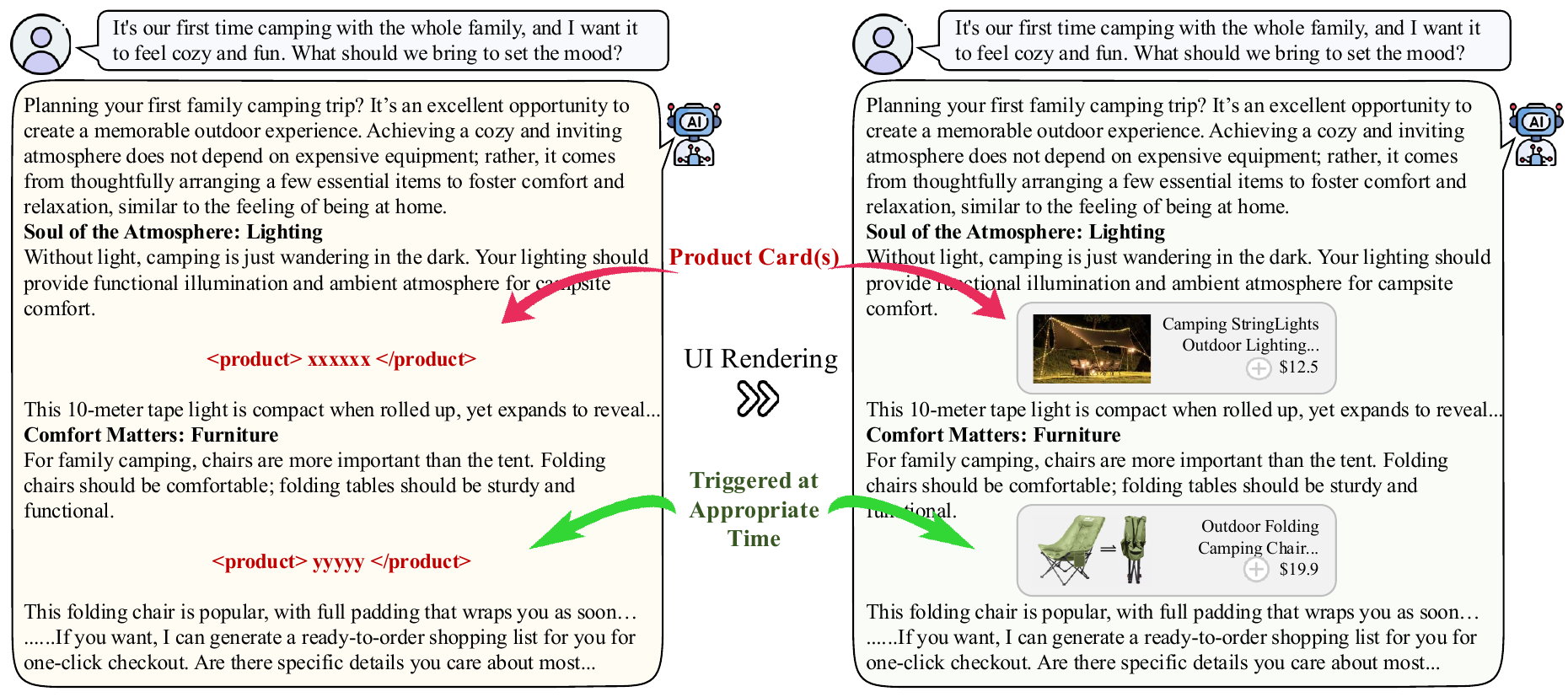}
    \caption{Overview of our conversational shopping agent.}
    \label{fig:product}
\end{figure*}

\textbf{Product Correctness} ensures that the recommended products not only satisfy the user’s requirements but are also presented accurately and consistently within the expected UI format. This dimension involves four complementary checks that jointly verify relevance, proper display, appropriate triggering, and completeness of the product recommendations, as illustrated in Figure~\ref{fig:product}:
\begin{enumerate}[leftmargin=3em]
\item UI Trigger:  Cards must be shown at the appropriate time.
\item UI Format: Cards must follow the required structure.
\item UI Relevance:  Products must meet the user's core needs.
\item UI Completeness: Mentioned products must have corresponding explanations.
\end{enumerate}

\textbf{Description Faithfulness} verifies that textual product descriptions accurately reflect actual product attributes. The LLM judge detects hallucinated features or misleading statements that could undermine user trust.

\textbf{Text Relevance} measures whether the textual response directly addresses the user query. The LLM judge identifies any off-topic or tangential content and verifies that the response provides coherent guidance relevant to the user’s request.

\subsection{L2 Grader}
\label{appendix:l2_grader}
The L2 Grader assesses the structural coherence and content depth of responses. Evaluation is conducted via a rubric-based assessment across seven metrics organized into two core dimensions:

\textbf{Structure:} Evaluates the logical framework and determines if the response follows a systematic decision-making process.
\begin{enumerate}[leftmargin=3em]
\item Core Decision Axis: Clearly identifies the key factors driving the decision and explains their causal roles.
\item Logical Consistency: Ensures coherence throughout the response.
\item Actionable Next Step: Provides concrete guidance to support user decision-making.
\end{enumerate}

\textbf{Depth:} Assesses the level of comparative insight, strategic prioritization, and risk awareness provided.
\begin{enumerate}[leftmargin=3em]
\item Path Differentiation: Explains differences between alternative choices or outcomes.
\item Route Prioritization: Justifies why one option or category is preferred over others.
\item Product-Level Comparison: Explains why a specific product is selected relative to alternatives.
\item Risk Mitigation: Identifies potential risks, trade-offs, or suboptimal choices.
\end{enumerate}

The prompt of L2 Grader is in Figure~\ref{l2_prompt}.

\section{Implementation Details}
\label{appendix:implementation_details}
We use Qwen3-30B-A3B-Thinking-2507~\cite{qwen3_tech_report} as the base model. The agent is first supervised fine-tuned on 480 high-quality trajectories annotated by DeepSeek-V3.2-reasoner~\cite{deepseekv32}, with batch size 8 and learning rate 7.0e-6 for 2 epochs, then trained with DCPO for 2 epochs using batch size 16 and learning rate 1e-6. During RL training, we sample $K=16$ trajectories per query and select $K/2=8$ via dynamic contrastive selection. Maximum sequence length during training is 32,768 tokens.
For inference, we use maximum context length 81,920 tokens, temperature 1.0, and top-p 0.9. 

For reward computation, both the L1 and L2 graders, as well as the process reward, are based on DeepSeek-V3.2-reasoner~\cite{deepseekv32}. We set $\alpha=0.5$, $\beta=0.05$, $\eta=0.7$, $k=5$ in HRM. 
For the tool implementations, \textsc{web\_search} uses Zhipu's \texttt{search\_pro} API and \textsc{product\_search} is implemented based on our commercial product search API from real-world industrial deployment.
In our training pipeline, supervised fine-tuning is conducted using LLaMA Factory~\cite{zheng2024llamafactory}. 
RL is performed with the VeRL~\cite{verl} framework, enabling multi-node distributed training.

\section{Analysis of Human Calibration}
To validate the reliability of the LLM-based grader, we conduct human calibration by comparing LLM judgments with human annotations on a shared evaluation subset. Human annotators evaluate responses using the same rubric as the LLM grader.

For L1 Grader evaluation, we measure agreement between human and LLM judgments using Cohen’s Kappa. For L2 Grader evaluation, we measure alignment using Spearman correlation between human and LLM scores across the seven evaluation dimensions. 

The results show strong alignment between the LLM grader and human judgments for both L1 and L2 Grader. The final agreement reaches $\kappa=0.79$ for L1 Grader and $\rho=0.71$ for L2 Grader. Overall, the calibration results support the use of the LLM grader as a reliable proxy for human evaluation.

\begin{figure*}[htbp]
\centering
\begin{tcolorbox}[
    enhanced jigsaw,
    breakable,
    pad at break*=1mm,
    colframe = gray!115,       % 边框颜色
    colback = gray!5!white,             % 背景颜色
    coltitle = white,                   % 标题字体颜色
    coltext = black,                    % 文字颜色
    % vfill before first,
    % break at=0pt/0pt,
    % %enlarge page flexible,
    % lines before break=4,
    % height fill=false,
    fonttitle = \bfseries,              % 标题字体加粗
    title = The Prompt of L1 Grader,  % 标题内容
    boxrule = 1pt,                      % 边框宽度
    arc = 1.5mm,                          % 边角圆润
    width = \textwidth,                 % 宽度
    left = 7pt,                         % 左边距
    right = 7pt,                        % 右边距
    top = 5pt,                          % 上边距
    bottom = 5pt                       % 下边距  
]
\fontsize{8.5pt}{10pt}\selectfont
You are a professional evaluation expert responsible for assessing the performance of an AI Agent. You will receive:
\begin{itemize}
    \item The user’s original query
    \item The Agent’s complete execution trajectory (including all tool calls and results)
    \item The Agent’s final response (including text and UI cards)
\end{itemize}
Your task is to evaluate the Agent’s performance across three dimensions.

\textbf{UI Card Format Description}

The UI card format is: \texttt{<tag>id(s)</tag>}

Available card types:
\begin{itemize}
    \item \texttt{<product>}: E-commerce product (ID starts with PD\_)
\end{itemize}
Multiple product IDs can be combined with commas to represent a bundle card recommendation:
\begin{itemize}
    \item Example: \texttt{<product>PD\_123,PD\_456</product>} means two products are displayed in a single card.
\end{itemize}

\textbf{Evaluation Criteria}

\begin{enumerate}[leftmargin=1em]
\item \textbf{Description Faithfulness}

[The criteria of Description Faithfulness]

\item \textbf{UI Completeness (UI Card Completeness and Timing)}

[The criteria of UI Completeness]
\item \textbf{Text Relevance}

[The criteria of Text Relevance]

\end{enumerate}

\textbf{Input Content:}  
\begin{itemize}
    \item \textbf{User Query:} \{query\} 
    \item \textbf{Agent Execution Trajectory:} \{trajectory\}
\end{itemize}

\textbf{Output Format Requirement:} The Agent’s evaluation must be a valid JSON object as follows:
\begin{alltt}
\{
  "description_faithfulness": {
    "is_pass": true/false,
    "reason": "Provide detailed reasoning supporting your evaluation"
  },
  "ui_completeness": {
    "is_pass": true/false,
    "reason": "Provide detailed reasoning supporting your evaluation"
  },
  "text_relevance": {
    "is_pass": true/false,
    "reason": "Provide detailed reasoning supporting your evaluation"
  }
\}
\end{alltt}
\textbf{Important Notes:}
\begin{itemize}
    \item Include specific examples in the reason fields to support your evaluation
    \item Stay objective and focus on concrete issues
    \item Distinguish between minor issues and serious failures
    \item Output only JSON; no extra text outside the JSON structure
\end{itemize}
\end{tcolorbox}
\caption{The prompt of L1 Grader.}
\label{l1_prompt}
\end{figure*}

% \begin{figure*}[htbp]
% \setlength{\abovecaptionskip}{0.2cm}
% \setlength{\belowcaptionskip}{0.2cm}
%\clearpage
\begin{figure*}[htbp]
\centering
\begin{tcolorbox}[
    enhanced jigsaw,
    breakable,
    pad at break*=1mm,
    colframe = gray!115,       % 边框颜色
    colback = gray!5!white,             % 背景颜色
    coltitle = white,                   % 标题字体颜色
    coltext = black,                    % 文字颜色
    fonttitle = \bfseries,              % 标题字体加粗
    title = The Prompt of L2 Grader,  % 标题内容
    boxrule = 1pt,                      % 边框宽度
    arc = 1.5mm,                          % 边角圆润
    width = \textwidth,                 % 宽度
    left = 7pt,                         % 左边距
    right = 7pt,                        % 右边距
    top = 5pt,                          % 上边距
    bottom = 5pt                       % 下边距  
]
\fontsize{8.5pt}{10pt}\selectfont
You are an evaluation expert for e-commerce agents, responsible for rigorously and objectively assessing the quality of an Agent’s response based on the given Rubrics.  
Your task is \textbf{not} to give an overall score or summary, but to: \textbf{evaluate whether the Agent’s response satisfies each Rubric individually, one by one.}

\textbf{Evaluation Principles (Must Follow):}

\begin{enumerate}[leftmargin=1em]
\item \textbf{Independent evaluation per Rubric:} 
\begin{itemize}
    \item Each Rubric must be evaluated individually. 
    \item The evaluation of the current Rubric must not be affected by whether other Rubrics have passed or failed.
\end{itemize}

\item \textbf{Strictly follow the literal meaning of the Rubric:} 
\begin{itemize}
    \item Judge only the content explicitly required by the Rubric.  
    \item Do not assume or add abilities that the Rubric does not explicitly require.  
    \item If a Rubric requirement is not clearly satisfied, it must be marked as not passed.
\end{itemize}

\item \textbf{Evidence-based judgment:} Evaluation can only be based on:
\begin{itemize}
    \item User Query
    \item Agent response text
    \item Any UI cards included in the Agent response and their official information
\end{itemize}
Do not introduce external knowledge or personal preference.
\item \textbf{UI card usage rules:} 
\begin{itemize}
    \item If a Rubric involves specifications, price, brand, target users, or other factual details, prioritize the official information in the UI cards.
    \item If the Agent’s text conflicts with the UI cards, use the UI card information as the source of truth and mark the Rubric as not passed.
\end{itemize}

\item \textbf{Strong vs. weak evidence:}
Many Rubrics evaluate whether a decision logic is explicitly adopted, not just mentioned. When making judgments, this distinction must be clear:  

\textit{Strong evidence (passable):}  
The relevant content is clearly, centrally, and perceptibly expressed, typically shown as:  
\begin{itemize}
    \item Explicitly stating a selection principle, decision axis, or trade-off before recommendations or in a summary
    \item Clearly saying ``I will prioritize/focus/first consider …''
    \item Developing this point as an independent evaluation dimension (at least one or two clear sentences)
\end{itemize}

\textit{Weak evidence (must fail):}  
The relevant content appears only in the following forms:  
\begin{itemize}
    \item Mentioned incidentally in a single product description (e.g., ``non-toxic'' as a one-line selling point)  
    \item Appears only in UI labels, review tags, or title keywords, without being explained as a decision logic  
    \item Only vague hints or scattered keywords, making it difficult for the user to perceive it as a main selection criterion
\end{itemize}
If a Rubric requires ``explicitly stated / clearly proposed / as a principle,'' but the evidence is only weak, it must be marked as not passed.

\item \textbf{Examples constraint:}  
Each Rubric may include \texttt{pass\_examples} (ideal structure) and/or \texttt{fail\_examples} (poor structure):  
\begin{itemize}
    \item If the Agent’s behavior is closer to the \texttt{fail\_examples} (poor structure), it should be marked as not passed.  
    \item Only if the Agent’s behavior is clearly closer to the \texttt{pass\_examples} (ideal structure) should it be marked as passed.   
    \item If a Rubric does not provide examples, it must be judged strictly according to the literal meaning of the Rubric.
\end{itemize}
\end{enumerate}

\textbf{Input Content:}  
\begin{itemize}
    \item \textbf{User Query:} \{query\} 
    \item \textbf{Agent Response:} \{response\}
    \item \textbf{UI Cards in the Agent Response (XML tags) and their official information:} \{ui\_card\_info\}
    \item \textbf{Rubrics (structured JSON, evaluate in order; do not change the order):} \{rubrics\_json\}
    \item \textbf{Structure Example:} \{structure\_whole\_example\}
\end{itemize}

\textbf{Output Requirements (Strict):}  
Please strictly output the evaluation results in the following JSON array format:
\begin{itemize}
    \item The length of the array must exactly match the number of Rubrics.  
    \item The N-th object in the array corresponds to the evaluation result of the N-th Rubric.  
    \item Do not add, remove, merge, or reorder Rubrics.
\end{itemize}

Each JSON item format:
\begin{alltt}
\{
  "is_pass": true or false,
  "reason": "Explaining why it does or does not satisfy the Rubric"
\}
\end{alltt}

Now, based on the rules above, evaluate each Rubric and output the results.
\end{tcolorbox}
\caption{The prompt of L2 Grader.}
\label{l2_prompt}
\end{figure*}

\end{document}